\documentstyle{article}

\begin{document}

\title{QCD sum rules as a tool for investigation of the baryon
properties at finite densities }
\author{E.G. Drukarev  and M.G. Ryskin\\
Petersburg Nuclear Physics Institute\\
Gatchina, St.Petersburg 188350, Russia }
\date{} \maketitle
\underline{\bf Introduction. } Speaking about the properties of nucleons
in nuclear matter, we have in mind, e.g.:
\begin{enumerate}
\item Potential energy of a nucleon in the medium.
\item Neutron-proton mass splitting in isotope-symmetric matter.
\item Parameters which describe  the interaction of nucleons with
external fields: axial coupling constant $g_A$ and magnetic moments
$\mu_N$.
\item Structure functions of deep inelastic scattering.\\ Turning to the
strange baryons we can add.
\item Properties of a strange baryon in nuclear matter.
\item The system of strange baryons ("strange matter").
\end{enumerate}
Approach of traditional nuclear physics to description of properties of
baryons in nuclear matter is based on conception of $NN$ interactions.
The problem is that small internucleon distances, where the nucleons
cannot be considered as structureless point particles appear to be of
crucial importance. Thus the whole approach becomes complicated and not
well defined.

However, while $NN$ interactions are complicated at small distances,
the strong interactions are not. Indeed, due to asymptotic freedom of
QCD, the latter are the perturbative interactions between quarks and
gluons. The peculiarity of QCD is the finite value of the vacuum
condensates of quark and gluon fields $\langle0|\bar qq|0\rangle$,
$\langle0|\frac{\alpha_s}\pi G^2|0\rangle$, etc. This means that in the
ground state of QCD there are finite densities of quark--\-antiquark
and gluon fields.

Shifman et al. \cite{1} suggested the QCD sum rules (SR) for the
description of characteristics of free mesons. The method was based on
the features of QCD, mentioned above. Later it was expanded by Ioffe
\cite{2} to the case of free baryons. Characteristics of free nucleons
where expressed through the values of QCD condensates.

In 1988  Drukarev and Levin \cite{3} used the SR method for
investigation of the properties of nucleons in nuclear matter. In
\cite{3} the first steps were made to express the potential energy of
the nucleon through in-\-medium values of QCD condensates. This paper
was followed by a number of works of Petersburg (Leningrad) Nuclear
Physics Institute --- (PNPI) group \cite{4}--\cite{9}. In 1991 the
Maryland University group joined this field of investigations
\cite{10}. Also a number of papers on meson properties in nuclear
matter was published later.\\

\underline{\bf QCD sum rules in vacuum \cite{1,2}. } They are based on
dispersion relation
\begin{equation}
\Pi_0(q^2)\ =\ \frac1\pi\int \frac{\mbox{Im }\Pi_0(k^2)dk^2}{k^2-q^2}
\end{equation}
for the function $\Pi_0(q^2)$ which describes the propagation of the
system carrying the quantum numbers of the nucleon (proton). Equation
(1) is considered at $q^2\to-\infty$ where the system can be treated
just as three quarks with perturbative  interactions between themselves
and with quarks and gluons of vacuum. At $q^2\to-\infty$ $\Pi_0(q^2)$
can be presented as power series
\begin{equation}
\Pi_0(q^2)\ =\ \sum^2_{n=0}a_nq^{2n}\ln q^2+\sum^\infty_{n=0} c_nq^{-2n}
\end{equation}
 known as operator
expansion. The coefficients $a_n,c_n$ are related to expectation values
of certain QCD operators. As to the right-hand side (r.h.s.) of Eq.(1),
the spectral density Im$\,\Pi_0(k^2)$ is related to observable spectrum
of the system. The usual approach is to single out the lowest laying
state (proton), approximating the higher states by continuum:
\begin{equation}
\mbox{Im }\Pi_0(k^2)\ =\ \lambda^2\delta(k^2-m^2)+\theta(k^2-W^2)\Delta
\Pi_0(-k^2)\ .
\end{equation}
This is known as "pole + continuum" model. Parameters $m$ and
$\lambda^2$ which describe the position of the lowest laying pole and
the residue are characteristics of proton. Continuum threshold $W^2$ is
the parameter of the model: the cut with physical threshold and unknown
spectral density is replaced by that with unknown value of the
threshold $W^2$ and fixed spectral density $\Delta\Pi_0(-k^2)$. The
special mathematical ansatz, the Borel transform (inversed Laplace
transform) increases the role of lower laying states. A function of
$q^2$ transforms into the one of Borel mass $M^2$, e.g.
\begin{equation}
\widehat B\ \frac1{k^2+q^2}\ =\ \exp\left(-\frac{k^2}{M^2}\right)\ .
\end{equation}

The model for the left-hand side (l.h.s.) of Eq.(1) becomes
increasingly true at large values of $M^2$. The one for r.h.s. works
better at small $M^2$. The basic assumption of the method is that there
is certain region of the values of $M^2$ in which both r.h.s. and l.h.s. of
Eq.(1) approximate the true function $\Pi_0(q^2)$ well enough. Then the
parameters $m,\lambda^2$ and $W^2$ can be expressed through the
values of QCD condensates. Ioffe \cite{2} found that the value of $m$
depends mainly on the condensate $\langle0|\bar qq|0\rangle$.

Thus the picture of formation of the proton mass turned out to be very
simple. It appears due to the exchange by quarks between our probe
system and the quark-\-antiquark pairs of QCD vacuum.

\underline{\bf QCD sum rules in nuclear matter. Calculation of
potential energy \cite{3,4,6}.} The generalization of the SR method to
the case of finite densities is not straightforward. The spectrum of
the function $\Pi(q)$ is more complicated now. One should single-out
the singularities connected with the baryon but not with the medium
itself. This can be done by the special choice of variables. Neglecting
the Fermi motion of the nucleons of the matter, we can fix the pair
energies $S$ of our probe hadron and that of the matter. Presenting the
QCD SR for the function $\Pi(q)=\Pi(q^2,s)$ we can single-out the
singularities connected with the probe hadron until we limit ourselves
to its pair interactions with the nucleons of the matter. In this
approach "pole+continuum" model, employed for vacuum can be used. This
choice of variables insures the condition $q_0\to\infty$ at
$q^2\to-\infty$ which is necessary for the operator expansion of the
function $\Pi(q)$. Another problem comes since each term of operator
expansion corresponds, in the general case, to infinite number of
condensates. Due to the presence of  logarithmic loops, several lowest
order terms of operator expansion contain, however, finite number of
the condensates.

In the papers \cite{3,4,6} QCD SR in nuclear matter were presented as
Borel transformed dispersion relations for the difference of the
functions $\Pi(q)$ in matter and in vacuum. The shifts of the
parameters $m,\lambda^2$ and $W^2$ caused by interaction with the
matter $(\Delta\,m$, $\Delta\lambda^2$ and $\Delta W^2$) were expressed
through in-medium values of QCD condensates.

On the other hand the shift of the position of the nucleon pole in
external field is $\Delta m=U$ with $U$ standing for the potential
energy of the nucleon. It was found that the value of $U$ is determined
mainly by the averaged values of the quark operators $\bar q\gamma_0q$
and $\bar qq$. The condensate $\langle M|\bar q\gamma_0q|M\rangle$,
which vanishes in vacuum, is just the density  of baryon number in the
system. The expectation value $\langle M|\bar qq|M\rangle$ is the
density of quark-antiquark pairs. Thus we come to a simple picture of
formation of the potential energy $U$. It comes from exchange by quarks
between our probe three quark system and the matter. The  latter can
contribute by its valence quarks $\langle M|\bar q\gamma_0q|M\rangle$
and by modification of its sea of quark-\-antiquark pairs $\langle
M|\bar qq|M\rangle-\langle0|\bar qq|0\rangle$.

One can immediately calculate the condensate
\begin{equation}
\langle M|\bar q\gamma_0q|M\rangle\ =\ \sum_i n_{q_i}\rho_i
\end{equation}
with $n_q$ being the number of $q$ quarks in a nucleon of the matter
($i$ denotes proton or neutron), $\rho_i$ stands for the density. The
SR analysis  provides the contribution $\Delta_vm\approx+200$ MeV
caused by this condensate (at $\rho_n=\rho_p=\rho/2$, $\rho=0.17$
Fm$^{-3}$). The scalar condensate can be presented as
\begin{equation}
\langle M|\bar qq|M\rangle-\langle0|\bar qq|0\rangle\ =\ \rho\langle
N|\bar qq|N\rangle+F(\rho)
\end{equation}
with the first term in r.h.s. of Eq.(6) standing for the gas approximation
while $F(\rho)$ describes the contribution of the meson cloud.
Fortunately the first term can be expressed through observables since
\begin{equation}
\langle N|\bar qq|N\rangle\ =\ \frac{2\sigma}{m_u+m_d}\ .
\end{equation}
Here $\sigma$ denotes pion-nucleon sigma term which can be extracted
from experimental data on low energy $\pi N$ scattering. The gas
approximation provides the contribution $\Delta_sm\approx-300$ MeV to
the potential energy. Several steps beyond the gas approximation have
been made also. The function $F(\rho)$ is determined mainly by the
relatively large distances of the order of inversed Fermi momenta or
larger ones. This makes some approximate calculations available. They
lead to the saturation curve with reasonable values of the equilibrium
density and of the binding energy.

\underline{\bf Other applications of the method. } One can generalize
the approach for the case of neutron-\-proton mass splitting in
symmetric matter. In QCD language this effect is caused by finite
values of the difference of quark masses $m_d-m_u$ and by the
non-\-vanishing value of the operator $\bar dd-\bar uu$. Both
contributions were included explicitly into QCD SR analysis \cite{7}.
Neutron was found to be bound stronger than the proton with reasonable
value of the mass difference.

As to parameters of interaction of the nucleons with external fields,
the application of the method at finite densities is the
straightforward generalization of this method in vacuum \cite{11}. In
the left-hand side of SR the quark system interacts with external field
while in the right-hand side the corresponding  parameter of nucleon
enters the equation. The first approach  to the calculation of
renormalization of axial coupling constant was made in \cite{5}.

In the same way the method was applied to the calculation of the deep
inelastic structure functions of nuclei. In this case the system
interacts with the hard virtual photon. In our paper \cite{9} we
calculated the deviations of the structure function $F_2$ from that of
a system of free nucleons at intermediate values of Bjorken variable
$x$. The calculated values followed typical EMC behaviour. As the next
steps of application of the approach to this problem we plan to
investigate cumulative aspects of the process. The method can be
applied also to investigation of gluon structure function. Another
interesting object is the structure function of a polarized nucleon.

One can see that the method can be applied for description of a strange
baryon in the matter. All the problems considered in this section and
in the previous one can be approached in the same way. Also behaviour
of a baryon in the system of strange ones can be described in terms of
the condensates of $u,d$ and $s$ quarks.

\underline{\bf Summary. } We made first steps in solving the problem of
expressing the characteristics of baryons at finite densities through
the in-medium values of the condensates. Potential energy of a nucleon
in nuclear matter was expressed as the sum of the terms proportional to
vector and scalar condensates. The former is  positive while the latter
is negative. Hence, the structure of potential energy reproduces that
of quantum hadrodynamics. The saturation of the matter in our approach
is provided by non-\-linear contribution to the scalar  condensate
$\langle M|\bar qq|M\rangle$. We obtained at least qualitative
description of neutron-\-proton mass splitting in nuclear matter. We
described also the influence of medium on nucleon structure functions.

Note that this approach does not describe quark effects only. It
describes the hadron effects, expressing them through certain quark
effects. For example exchange by mesons (pairs of strongly correlated
quarks) in the r.h.s. of the sum rules is expressed through exchange by
pairs of uncorrelated quarks in l.h.s.

We obtained some new knowledge. We show the scalar forces to be related
to $\pi N$ sigma term. In the case of isotope-\-breaking forces we show
the scalar channel to be as important as the vector one. Thus, the
method provides guide-\-likes for traditional nuclear physics.

All the results, described above, were obtained without fitting
parameters.   We did not use a controversial conception of $NN$
interaction.

Note  one more point. The QCD SR method provides a unique approach to
the problems, listed in the beginning of this paper. In framework of
traditional nuclear physics they require different knowledge and
different skill. Thus, usually they attract attention of different
communities of the explorers.

The method should be improved by inclusion of more complicated
in-\-medium condensates. Also the role of higher order terms of
expansion in powers of Fermi momentum should be clarified.

Note that investigation of QCD SR stimulated other directions of
research. Say, the first analysis of the function $\langle M|\bar
qq|M\rangle$ carried out in \cite{3} was followed by more than a dozen
works on the subject.

One of us (E.G.D.) is indebted to the Organizing Committee of the
Conference and to the Russian Fund for Fundamental Research (grant
\#97-02-27083) for the support.
This activity is supported by the Russian Fund for Fundamental
Research (grant\#95-02-03752-a).

\end{document}